\documentclass[%
 reprint,
 amsmath,amssymb,
 aps,
]{revtex4-1}
\newcommand{\be}{\begin{equation}}
\newcommand{\ee}{\end{equation}}

\usepackage{graphicx}
\usepackage{dcolumn}
\usepackage{bm}
\usepackage{color}

\begin{document}

\preprint{}

\title{Escape of cosmic rays from the Galaxy and effects on the circumgalactic medium }

\author{Pasquale Blasi$^{1,2}$ \& Elena Amato$^{3,4}$}
\affiliation{$^1$Gran Sasso Science Institute, Viale F. Crispi 7, 67100 L'Aquila, Italy\\
$^2$INFN/Laboratori Nazionali del Gran Sasso, Via G. Acitelli 22, Assergi (AQ), Italy\\
$^3$INAF-Osservatorio Astrofisico di Arcetri, Largo E. Fermi 5, 50125 Firenze, Italy\\
$^4$Dipartimento di Fisica e Astronomia, Universit\`a degli Studi di Firenze, Via Sansone 1, 50019 Sesto Fiorentino (FI), Italy}





\date{\today}

\begin{abstract}
	The escape of cosmic rays from the Galaxy is expected to shape their spectrum inside the Galaxy. Yet, this phenomenon is very poorly understood and, in the absence of a physical description, it is usually modelled as free escape from a given boundary, typically located at a few kpc distance from the Galactic disc. We show that the assumption of free escape leads to the conclusion that the cosmic ray current propagating in the circumgalactic medium is responsible for a non resonant cosmic ray induced instability that in turn leads to the generation of a magnetic field of strength $\sim 2\times 10^{-8}$ Gauss on a scale $\sim 10$ kpc around our Galaxy. The self-generated diffusion produces large gradients in the particle pressure that induce a displacement of the intergalactic medium with velocity $\sim 10-100$ km/s. Cosmic rays are then carried away by advection. If the overdensity of the intergalactic gas in a region of size $\sim 10$ kpc around our Galaxy is $\gtrsim 100$ with respect to the cosmological baryon density $\Omega_{b}\rho_{cr}$, then the flux of high energy neutrinos as due to pion production becomes comparable with the flux of astrophysical neutrinos recently measured by IceCube. 
\end{abstract}

\pacs{Valid PACS appear here}
\maketitle

{\it Introduction} - The spectrum of cosmic rays (CRs) in the Galaxy is shaped by diffusion, advection and the escape of particles from the confinement volume. The latter is typically modelled by assuming that particles reaching the halo boundary ($|z|=H$) can freely escape. For particles with energy above $\sim 1-10$ GeV/n, where energy losses are weak and transport should be dominated by diffusion, escape occurs on a time-scale $\tau_d\sim H^2/D(E)$, where $D(E)$ is the energy dependent diffusion coefficient in the confinement volume \citep{crbible}. The free escape boundary condition results in a CR gradient $\nabla n\sim n_{gal}(E)/H$, where $n_{gal}$ is the CR spectrum in the disc of the Galaxy. Outside the halo the particle motion is assumed to become ballistic. Then one can deduce the CR density outside the halo, $n_{out}$, from flux conservation, $D(E)\nabla n=(c/3) n_{out}(E)$, finding $n_{out}\simeq \frac{3D(E)}{H c} n_{gal}(E)$. Recalling that the diffusion coefficient can be written in terms of the diffusion pathlength $\lambda_{D}$ as $D(E)=(1/3) v \lambda_{D}$, and keeping in mind that for diffusing CRs $\lambda_{D}\ll H$, one infers that $n_{out}\ll n_{gal}$.  It is also worth noticing that, as long as energy losses do not play a crucial role, which is typically the case for the dominant proton component at energies above $\sim$ few GeV, the spectrum of CRs escaping the Galaxy is the same as the injection spectrum. These simple considerations remain valid even in models in which the diffusive confinement of CRs in the Galaxy is self-induced through the excitation of resonant streaming instability, as first suggested in \cite{Wentzel69,Kulsrud69,Jokipii76,holmes75}. Although the number density in the form of escaping particles is very small, the current they transport is the same as inside the Galaxy, so that the question arises of whether such current can affect the medium in which CRs propagate outside the Galaxy.

In \cite{Dangelo} it was shown that the current of ultra high energy CRs (UHECRs) escaping sources such as Active Galactic Nuclei may generate large scale magnetic fields in the near source regions, thanks to the non-resonant streaming instability. Confinement of lower energy particles within these regions for the age of the Universe induces a low energy cutoff ($\lesssim 10^{7}-10^{8}$ GeV) in the spectrum of UHECRs that reach the Earth from such sources.

Here we investigate this phenomenon for our own Galaxy and its implications for what we call escape of CRs. We envision the escape as transport in a background magnetic field that decreases with distance from the disc while the diffusion coefficient rises. At some point particle transport may become ballistic. We find that as soon as the background field drops below some critical value, the current of escaping particles is sufficient to excite a non-resonant streaming instability \citep{bell2004} responsible for a rapid growth of magnetic field and effective particle scattering. This leads to a strong coupling between particles and background plasma. The latter is set in motion with a speed that is about the Alfv\'en speed in the amplified field, as was also found in numerical simulations of the saturation of the instability \cite{riquelme}. This phenomenon makes the transport of CRs dominated by advection rather than by diffusion, but it also implies that the free escape of CRs (at the speed of light) is rapidly slowed down to about the local Alfv\'en speed, thereby forcing us to reconsider the physical meaning of free escape. 

The occasional interactions of CRs advected away from the Galaxy with gas in the circumgalactic medium are responsible for the production of gamma rays and neutrinos with a quasi-isotropic pattern and a spectrum $\sim E^{-2}$.  This emission is dominated by distances smaller than $\sim 10-20$ kpc from the Galaxy. If the density of the circumgalactic medium is of order $\sim 100 \Omega_{b}\rho_{cr}$, (where $\rho_{cr}=1.88 \times 10^{-29} h^2\ {\rm g}\ {\rm cm}^{-3}$ is the critical density of the Universe and $\Omega_b h^2 = 0.022$) the neutrino flux at the Earth is comparable with the flux of astrophysical diffuse neutrinos observed by IceCube \cite{icecube1,icecube2}.

{\it Streaming instability excited by escaping CRs} - If the role of losses is neglected, which is a good approximation for CR protons at energies above $\sim$ GeV, the equilibrium spectrum of CRs in the Galaxy can be related to the diffusion coefficient $D(E)$ and to the injection rate of CR particles in the Galaxy, $Q_{\rm CR}$, through the expression  
\be
n_{gal}(E) = \frac{Q_{CR}(E)}{2\pi R_{d}^{2}}\frac{H}{D(E)},
\ee
where $R_d$ is the radius of the Galactic disk. Here we assumed for simplicity that the sources of CRs are localized in an infinitely thin disc and that CRs can freely escape at the edge of the halo, located at an altitude  $|z|=H$, where $n_{CR}\to 0$. 

The CR density at position $z$ in the halo is then found to be:
\be
n_{CR}(z,E)=n_{gal}(E) \left(1-\frac{|z|}{H}\right),
\ee
which implies that CRs escape from the boundary of the halo with a flux
\be
\phi_{CR}(E) = - D(E) \frac{\partial n_{gal}}{\partial z} = D\frac{n_{gal}}{H}= \frac{L_{CR}}{2\pi R_{d}^{2}\Lambda}E^{-2}\ ,
\label{eq:phi}
\ee
where we have assumed that CRs are injected in the Galaxy with a luminosity $L_{CR}$ and a spectrum $\propto E^{-2}$ extending between $E_{min}$ and $E_{max}$, and $\Lambda=\ln(E_{max}/E_{min})$. Eq.~\ref{eq:phi} clearly shows that, as expected, the spectrum of escaping CRs is the same as the injected spectrum. The flux of CRs escaping the Galaxy is as given in Eq. \ref{eq:phi} independent of the details of transport in the Galaxy. For instance in more complex models such the one in Ref. \cite{Evoli18}, in which the halo arises from resonant self-generation and cascading, $H$ may in principle depend on energy. Yet the flux of CRs escaping into the surrounding medium is still the same as in Eq. \ref{eq:phi}. If, as commonly assumed, CRs propagate ballistically outside the Galactic halo and into the intergalactic medium, then their density immediately outside the halo boundary can be easily estimated from flux conservation as $n_{CR,ext}(E)=3\phi_{CR}/c$. For our purposes, however, the assumption of ballistic motion is not essential. In fact we focus on the current carried by CRs with energy $>E$, given by $J_{CR}=e E\phi_{CR}(E)$. As discussed in \cite{bell2004} a non-resonant instability is induced by this current provided the energy flux associated with the escaping particles is larger than $c$ times the magnetic energy density pre-existing the current:
\be
\frac{E^{2}\phi_{CR}}{c} > \frac{B_{0}^{2}}{4\pi},
\label{eq:nr}
\ee
where we assumed that a regular magnetic field of strength $B_{0}$ is present in the circumgalactic medium (CGM) around our Galaxy. The instability is excited on scales that are initially much smaller than the Larmor radius of the particles dominating the current, namely at wavenumber
\be
k_{max} = \frac{4\pi}{c B_{0}}J_{CR}=\frac{4\pi}{cB_0^2}\frac{E^2\phi_{CR}}{r_L(E)}\ ,
\ee
and with a growth rate $\gamma_{max}=k_{max}v_{A}$, where $v_{A}=B_{0}/\sqrt{4\pi \rho}$ is the Alfv\'en speed in the unperturbed field and the density $\rho$ is written as $\delta_{G} \Omega_{b}\rho_{cr}$.
The parameter $\delta_{G}\gtrsim1$ allows us to account for an overdensity of baryons around the Galaxy. 

The condition for the excitation of the non-resonant instability, Eq.~\ref{eq:nr}, translates into a condition on the background magnetic field
\be
B_{0}\leq B_{sat}\approx 2.2 \times 10^{-8}  L_{41}^{1/2} R_{10}^{-1}\ {\rm G}
\label{eq:Bsat}
\ee
where $L_{41}$ is the CR luminosity of the Galaxy in units of $10^{41}$ erg s$^{-1}$ and $R_{10}$ is the radius of the galactic disk in units of 10 kpc. We assumed $E_{max}=1 PeV$. However this parameter only enters all estimates logarithmically. It should also be noticed that even if the injection spectrum of CRs in the Galaxy were somewhat steeper, say $\propto E^{-2.2}$ rather than $\propto E^{-2}$, as some models suggest, the resulting lower flux of high energy particles would only decrease the magnetic field strength estimated in Eq.~\ref{eq:Bsat} by a factor $\sim 2$.

In order to check whether Eq.~\ref{eq:Bsat} is likely to be satisfied, one can estimate an upper limit on $B_0$ based on equipartition with the thermal energy density: for a CGM density $\delta_{G}\ \rho_{cr}\ \Omega_{b}$ and temperature $T$, this results in $B_{th}\leq 2 \times 10^{-9} {\rm G}\ \delta_{G}^{1/2} T_{4}^{1/2}$, where $T_{4}$ is the gas temperature in units of $10^{4}$ K. We expect $B_0$ well below $B_{th}$ in most cases, which suggests that it is safe to assume that the condition for the development of the non resonant instability is typically satisfied away from the Galaxy, where $\delta_{G}\lesssim$ a few hundreds. 

When the instability is excited, its growth proceeds at a rate
\be
\gamma_{max}=k_{max}v_A\approx 2\ {\rm yr}^{-1}\ \delta_G^{-1/2} E_{\rm GeV}^{-1} L_{41} R_{10}^{-2}\ .
\ee
For reasonable values of $\delta_G$ the time for growth is extremely short compared to all other relevant timescales, so that the field rapidly grows. 
It should be noticed that the initial growth rate of the instability might be modified by thermal effects in the situation we are considering. These effects become important when the IGM temperature and initial magneic field strength satisfy the condition \citep{reville08,zweibel10}: $B_*<1.8\times 10^{-9}\ \delta_G^{1/6}\ T_6^{1/3}\ (L_{41}/(E_{GeV} R_{10}^2) )^{1/3}$. In this case, the growth rate of the instability is reduced, but only by a factor $\sim 2$ for the values of the parameters typical of our problem. As soon as the field strength grows beyond $B_*$, we expect the evolution of the system to proceed as in the cold case.

The magnetic field growth initially happens on scales $k_{max}^{-1}$ much smaller than the Larmor radius of the particles dominating the current, so that the current is only weakly affected by the growth. On the other hand, at the same time a force $\sim J_{CR}\delta B/c$ is exerted on the background plasma, that gets displaced by an amount $\Delta r \sim \delta B J_{CR}/c \rho \gamma_{max}^{2}$. The instability eventually saturates when the scale $\Delta r$ becomes of the same order of magnitude of the Larmor radius, which implies
\be
\delta B\approx B_{\rm sat}\approx\sqrt{\frac{2 L_{CR}}{c\ R_d^2\Lambda}}
\label{eq:saturation}
\ee
It is important to notice that, for a spectrum $N(E)\propto E^{-2}$, $\delta B$ is the same on all scales, so that the diffusion coefficient is expected to be Bohm-like. Moreover $\delta B$ is independent of the initial magnetic field strength and the density of background plasma. 

The Bohm diffusion coefficient corresponding to this situation is 
\be
D(E) = \frac{1}{3}\frac{E\ c}{e\delta B} \approx 1.5\times 10^{24}\ E_{\rm GeV}\ L_{41}^{1/2} R_{10}\ \rm cm^{2} ~ s^{-1}\ .
\label{eq:Diff}
\ee
In other words, the original assumption of free streaming of CRs after escaping our Galaxy leads to the apparently contradicting result that the instability they excite is sufficient to induce a diffusive motion with short scattering length, hence particles diffuse very slowly as soon as they find themselves in a region where condition \ref{eq:Bsat} is satisfied. On the other hand this conclusion does not really depend on any specific assumption on the physics of particle propagation, while only based on conservation of the energy flux constantly injected in our Galaxy in the form of CRs.

Much discussion has appeared in the literature concerning the saturation of the instability. A comprehensive study of the topic \cite{riquelme} has highlighted two processes that may limit the saturation field to lower values than the one derived above. The first is the progressive increase with growing field strength of the fastest growing wavelength. We checked that this process leads to saturation values close to that in Eq.~\ref{eq:saturation}, although a dependence on $B_{0}$ appears. Another process turns out to be of greater importance: while the field grows and the motion of CRs from ballistic turns into diffusive, the pressure gradient that develops exerts a force on the background plasma. This force sets the gas in motion and the field saturates when the plasma bulk speed equals the Alfv\'en velocity in the amplified field, $\tilde v_{A}=\delta B/\sqrt{4\pi \rho}$. If diffusion were the dominant transport process for CRs outside the halo, the density could be approximated as
\be
n_{CR}(E) \approx 2\phi_{CR}\sqrt{\frac{t}{\pi D(E)}},~~~~~\rm z<\sqrt{4 D(E) t},
\ee
where we assumed that CRs enter steadily through the halo surface and diffuse with a constant diffusion coefficient $D(E)$ for a time $t$. Notice that despite the fact that diffusion acts in three spatial dimensions, on scales $z<R_{d}$ the solution of the diffusion equation retains its 1D scalings.

For the scales that are reached by particles in a time $t$, namely for $z<\sqrt{4 D(E) t}$, the current of particles is conserved and still given by Eq.~\ref{eq:phi}. Hence we can define a diffusive (or drift) velocity
\be
v_{D}=\frac{\phi_{CR}}{n_{CR}}\approx\sqrt{\frac{\pi D(E)}{4 t}}\ .
\label{eq:vd}
\ee
At the same time, diffusion creates a pressure gradient that is directly related to the CR current as 
\be
\nabla P_{CR}=E^2 \phi_{CR}/D.
\label{eq:crpress}
\ee
This force imparts to the background plasma a velocity that can be estimated as 
\be
v_{bg} \approx \frac{E^2 \phi_{CR}}{D}\frac{t}{\rho}\ .
\label{eq:vbg}
\ee
We argue that the chain of physical processes setting the saturation level of the magnetic field is then the following: the CR current produces an increase in the level of magnetic fluctuations which in turn reduce the CR diffusion coefficient (Eq.~\ref{eq:Diff}); this produces a CR pressure gradient (Eq.~\ref{eq:crpress}) which sets the background plasma into motion; a stationary situation is achieved when CRs and the plasma move at roughly the Alfv\`en speed in the amplified field: $v_D\approx v_{bg}\approx \tilde v_A = \delta B/\sqrt{4\pi \rho}$. Using the definitions of $v_D$ and $v_{bg}$ given in Eqs.~\ref{eq:vd} and \ref{eq:vbg} and the diffusion coefficient from Eq.~\ref{eq:Diff}, one derives: 
\be
\delta B \approx \left(\frac{\pi^{3/2} \rho^{1/2} L_{CR}}{R_d^2 \Lambda}\right)^{1/3}\approx 3\times10^{-7} L_{41}^{1/3} \delta_G^{1/6} R_{10}^{-2/3} {\rm G} .
\label{eq:dbadv}
\ee
This criterion returns then a strength of the magnetic field that is somewhat larger than in Eq.~\ref{eq:saturation}, for typical values of the parameters. Hence we argue that the field saturates at the level $\delta B\leq B_{sat}\approx 2.4 \times 10^{-8}$ G, as deduced from Eq.~\ref{eq:saturation}, without ever reaching the strength in Eq.~\ref{eq:dbadv}. Yet the advection velocity $\tilde v_{bg}$ becomes larger than $v_{D}$ after a time of a few $\gamma_{max}^{-1}$, indicating that the transport quickly turns into advective, so that the distribution function of CRs is 
\be
n_{CR}(E) = \frac{\phi_{CR}}{\tilde v_{A}}, 
\ee 
a factor $\approx c/\tilde v_A$ larger than what would be estimated in the case of ballistic motion at the speed of light.

The problem can be treated as one dimensional, as we have implicitly assumed above, for as long as $z\lesssim R_{d}$, which happens in a time $\sim R_{d}/\tilde v_{A}$. At larger distances the CR density drops with spherical radius $r$ as $r^{-2}$ and the effects described above quickly disappear. 
\vskip .1cm
{\it Implications for CR escape} - at first it may seem counterintuitive that the assumption of escape of CRs across a free escape boundary leads to the conclusion that in fact particles create an exceedingly small diffusion coefficient. The picture we have in mind is that of CRs propagating in an exponentially decreasing magnetic field while leaving the Galaxy \cite{ferriere}. This results in a rapidly growing diffusion coefficient. Effectively such a rapid growth of the scattering length is equivalent to a transition to a quasi-ballistic motion. At a distance where the energy density in magnetic fields drops below that of the escaping particles (Eq. \ref{eq:Bsat}), the CR current driven instability starts being excited and within a time of order several $\gamma_{max}^{-1}$ the field grows to $\sim 0.02 \mu$G, which implies a strong tie between CRs and background plasma. The pressure gradient induced by diffusion sets the plasma in motion and CR transport becomes advective with a typical velocity $\tilde v_{A}\sim 10-100$ km/s, depending on the gas density in the circumgalactic region. Advection carries CRs away from the Galaxy, inhibiting their return. Hence the equilibrium CR distribution inside the Galaxy (and more specifically in the disc) is not changed with respect to the standard picture we use to describe CR transport in the Galaxy. 
\vskip .1cm
{\it Implications for the magnetization of the Universe} - the escape of CRs from any galaxy leads to the creation of an extended region where the magnetic field is pushed up to a value close to the equipartition value with the current of escaping CRs. Each galaxy can be imagined as embedded in a halo of CR induced magnetic field. The extent and strength of such field depends on the type of galaxy and its luminosity. The magnetic field originated from the growth of a CR current is expected to be roughly scale invariant, which implies that the largest scale is of the order of the Larmor radius of the highest energy particles in the current. Largest scales are expected around galaxies that host more powerful accelerators, able to accelerate particles to higher energies. As discussed in \cite{Dangelo}, the field can be large enough to affect the transport of low energy CRs from the sources to Earth, by allowing only ultra high energy particles to reach us. 
\vskip .1cm
{\it Implications for high energy neutrinos} - the region of size $\sim R_{d}$ around our Galaxy is filled with CRs in a time $\sim R_{d}/\tilde v_{A}\sim 10^{8}-10^{9}$ years, which corresponds to a local overdensity of CRs in the same region of $\sim c/\tilde v_{A}\sim 10^{4}-10^{5}$ compared with the case of free escape. The occasional interactions of these particles with the CGM gas leads to production of gamma rays and neutrinos. The flux of neutrinos can be estimated as follows:
\be
F_{\nu}(E_{\nu}) E_{\nu}^{2} \approx  \frac{L_{CR}}{2 \pi R_d^2 \Lambda \tilde v_A}\frac{E_\nu^2}{E^2}
\frac{dE}{dE_{\nu}} \frac{\delta_{G} \Omega_{b} \rho_{cr}}{m_{p}} \frac{c \sigma_{pp} R_{d}}{2\pi}.
\ee

Calculating $\tilde v_A$ in the magnetic field $\delta B$ from Eq.~\ref{eq:saturation} we obtain
$$
F_{\nu}(E_{\nu}) E_{\nu}^{2} \approx \left(\frac{L_{CR}}{\Lambda}\right)^{\frac{1}{2}}\left(\frac{c \delta_{G}\Omega_{b}\rho_{cr}}{2\pi}\right)^{3/2}
\frac{\eta \sigma_{pp}}{m_p} = 
$$
\be
5\times 10^{-12} \delta_{G}^{3/2} \rm GeV cm^{-2} s^{-1} sr^{-1},
\ee
where we assumed that the neutrino energy is related to the energy of the parent proton by $E_{\nu}=\eta E$, with $\eta \sim 0.05$. For simplicity we neglected the weak energy dependence of the cross section for neutrino production, which is known to increase slowly with energy, so as to lead to a slight increase in the neutrino flux at high energy. This effect would partly compensate for a proton spectrum possibly steeper than $E^{-2}$ leading to a neutrino flux in any case flatter than the CR injection spectrum in the Galaxy and still compatible with the inferred slope of the IceCube neutrino spectrum, given the large error bars.

\begin{figure}
\begin{center}
\includegraphics[width=0.49\textwidth]{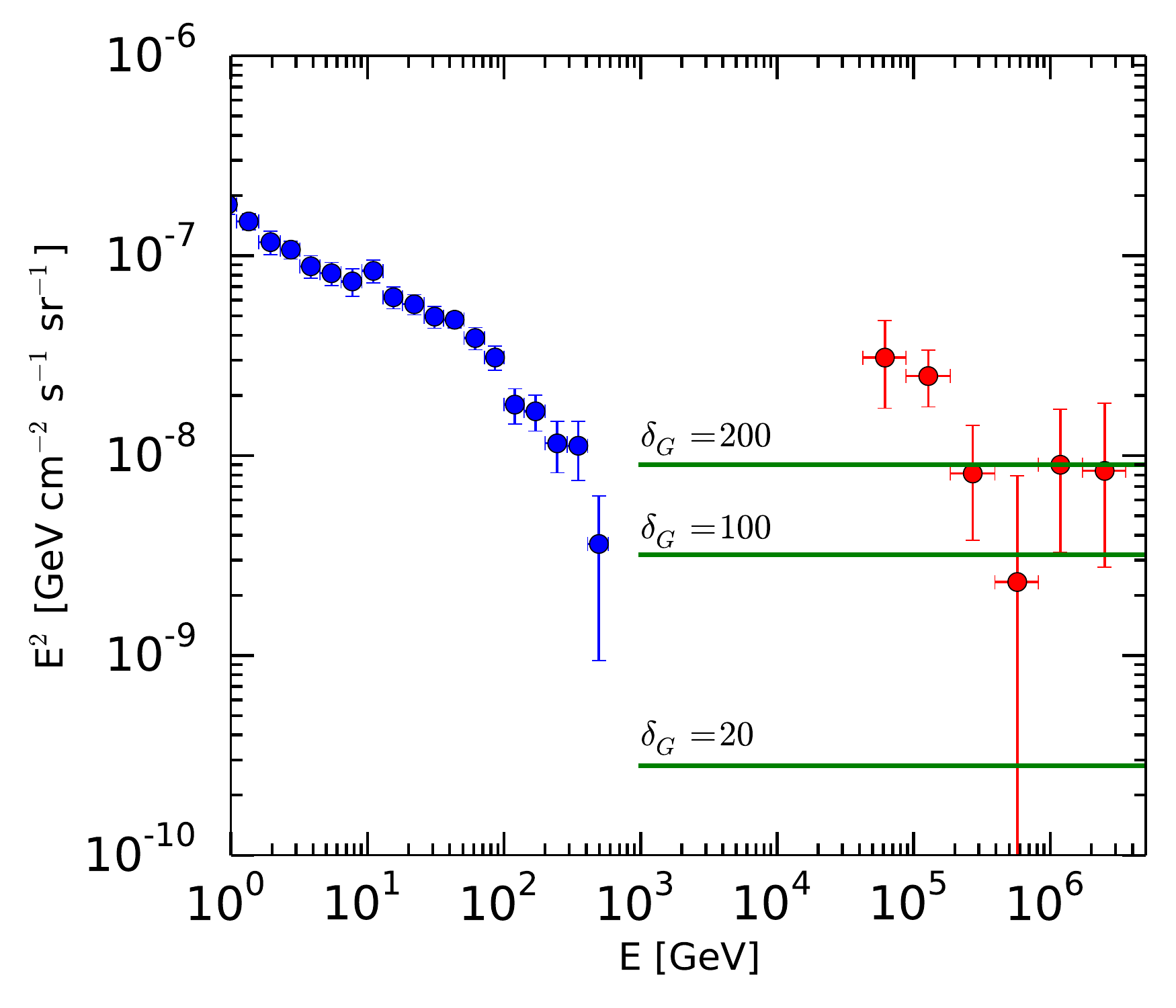}
\caption{Flux of isotropic diffuse gamma ray emission (blue) as measured by Fermi-LAT \cite{ackermann} and flux of astrophysical neutrinos as measured by IceCube \cite{icecube1,icecube2} (red). The (green) horizontal lines show the expected flux of neutrinos from pp collisions in the circumgalactic medium, for the overdensity $\delta_{G}$ as indicated.}
\label{fig:diffuse}
\end{center}
\end{figure}

The estimated flux of diffuse neutrinos is plotted in Fig. \ref{fig:diffuse} (green horizontal lines) for different values of the overdensity $\delta_{G}$. In the same figure we show the flux of astrophysical neutrinos measured by IceCube \cite{icecube1,icecube2} and, for comparison, the flux of gamma rays that Fermi-LAT associates with an isotropic extragalactic origin \cite{ackermann}. One can see that if the overdensity of baryonic gas in the circumgalactic medium is of order $\sim 100$, then the expected neutrino flux is comparable with the one measured by IceCube. It is worthwhile to mention that the virial radius of our Galaxy, which is of order $\sim 100$ kpc, is defined as the radius inside which the mean overdensity is 200. Hence a value of $\delta_{G}\sim 100-200$ appears to be quite well justified on scales of $\sim 10$ kpc. 

The production of neutrinos is also associated with the production of secondary electrons that reach equilibrium due to energy losses. We checked that the synchrotron radio emission of these electrons in the self-generated magnetic field outside the Galaxy is $10^{-4}-10^{-2}$ jy, depending on the value of $\delta_{G}$, several orders of magnitude smaller than the radio continuum emission of electrons in the halo.

A few caveats should be stressed: 1) the fluxes shown in Fig. \ref{fig:diffuse} have been obtained in the simple case that the injection spectrum from individual supernovae (or whatever other sources of Galactic CRs) is $N(E)\sim E^{-2}$. A steeper injection spectrum reflects in a correspondingly steeper spectrum of the neutrino flux. At the present level of investigation of the complex phenomenon of interaction of escaping CRs with the CGM, considering additional complications would shed no more light on the relevant physics. 2) In the estimate above we did not include a spectral suppression at some maximum energy. Such suppression is expected to reflect in a change of slope rather than a cutoff in the CR source spectrum \cite{schure,cardillo}, that would lead to a somewhat lower neutrino flux at the highest energies. 
\vskip .1cm
{\it Conclusions}: the escape of CRs from our Galaxy is required by the assumption that at energies $\gtrsim 10$ GeV stationarity is reached between injection at the sources and escape. This equilibrium is observationally visible in the decrease with energy of the ratio between the fluxes of secondary and primary nuclei, most notably the B/C ratio. Escape is not well understood and is usually modelled by imposing the existence of a free escape boundary located at a few kpc from the disc. At such boundary it is assumed that the CR transport becomes ballistic in nature. In reality we expect that the diffusion coefficient becomes increasingly larger away from the disc, so that at some point particles can be assumed to stream freely. We showed that, independently of the details of the propagation physics, the current of such escaping CRs in the circumgalactic medium is such that a rapidly growing, non resonant instability is excited as soon as the background magnetic field is lower than $\sim 2\times 10^{-8}$ Gauss. The growth of the instability leads to three main consequences: 1) a quasi-scale invariant magnetic field is generated with a strength $\sim 2\times 10^{-8}$ Gauss over a distance $\sim 10$ kpc from the Galaxy. 2) The pressure gradient of CRs scattering on such magnetic fluctuations sets the background plasma around our Galaxy in motion, so that CRs are advected away with the plasma at a speed $\sim 10-100$ km/s. 3) The occasional interactions of CRs with the CGM produce a quasi-isotropic neutrino flux at Earth, that is comparable with the flux observed by IceCube, provided the local baryon overdensity is $\sim 100$.  

{\it Acknowledgements---} We are grateful to Damiano Caprioli for interesting discussions on the saturation of the non resonant instability. 

\bibliography{escape} 

\end{document}